\documentclass[paper,twocolumn,showpacs,prl]{revtex4}%
\usepackage{amsfonts}
\usepackage{amsmath}
\usepackage{amssymb}
\usepackage{graphicx}%
\providecommand{\U}[1]{\protect \rule{.1in}{.1in}}

\begin{document}
\title{Direct optical detection of pure spin current in semiconductors}
\author{Jiang-Tao Liu}
\author{Kai Chang}
\altaffiliation[Corresponding author:]{kchang@red.semi.ac.cn}

\affiliation{SKLSM, Institute of Semiconductors, Chinese Academy of Sciences, P. O. Box
912, Beijing 100083, China}
\date{\today}


\begin{abstract}
We suggest a new practical scheme for the direct detection of pure spin
current by using the two-color Faraday rotation of optical quantum
interference process (QUIP) in a semiconductor system. We demonstrate
theoretically that the Faraday rotation of QUIP depends sensitively on the
spin orientation and wave vector of the carriers, and can be tuned by the
relative phase and the polarization direction of the $\omega$ and $2\omega$
laser beams. By adjusting these parameters, the magnitude and direction of the
spin current can be detected.

\end{abstract}
\maketitle


Generating spin population in semiconductors is one of central goals of
spintronics and has attracted a rapidly growing interest for its potential
application in spintronic devices. For pure spin current, the spin-up and
spin-down electron currents are expected to have equal magnitudes but travel
in opposite directions ($\pm \mathbf{k}$). In semiconductors, these currents
can be generated utilizing the spin Hall effect
(SHE),\cite{DP,Murak,Sinova,Kato} optical quantum mechanical interference
control between one- and two-photon
excitations,\cite{Bhat,Stevens,bner,hach,Yao} and one-photon absorption in the
noncentrosymmetric semiconductors.\cite{Bhat2,zhao} However, the experimental
measurement of spin current is an extremely challenging task. The pure spin
current can only be detected indirectly in semiconductors by measuring the
spin accumulation\cite{wunder,Bhat,bner,Kato,Stevens} near the boundary of the
sample, the nonuniform spatial distribution, i.e., spin wavepacket\cite{zhao}.
So far, the direct measurement of a pure spin current, with \emph{uniform}
spatial spin and charge distributions, has still not been reported in
semiconductor structures since the pure spin current exhibits vanishing charge
current and total spin. The vanishing charge current and total spin make
\emph{direct} electrical or optical detection in semiconductor structures
extremely difficult.

Is it possible to detect the \emph{uniform} pure spin current in a
semiconductor system? Although the total spin of a uniform pure spin
current vanishes everywhere in the real space, the spin population
is asymmetric in the momentum space, i.e., the asymmetric
distribution of the spin-up (-down) electrons at opposite $\pm
\mathbf{k}$ points. Direct optical detection would be possible if
the optical transition at $\pm \mathbf{k}$ points become asymmetric
utilizing a quantum interference process (QUIP), e.g., the
two-photon process. The momentum-resolved optical transition can be
detected by circularly polarized light, Faraday or Kerr rotation (FR
or KR), and magnetic circular dichroism (MCD). In this work, we
focus on the Faraday rotation spectrum because it is a powerful tool
for investigating the spin dynamics of carriers in semiconductors
\cite{kikk}. The conventional FR vanishes for the detection of pure
spin current since it is determined by the total spin of the system,
i.e., the difference of the spin-up and spin-down electron
populations.

In this Letter, we suggest a new scheme utilizing a momentum-resolved
two-color FR of optical quantum interference process for direct detection of
spin current in a semiconductor\emph{ }structure. In the quantum interference
process between one- and two- photon absorptions, the transition rate depends
not only on the polarizations of the laser pulses, but also on the the
relative phase of the $\omega$ and $2\omega$ excitations, which makes the
transition rate depending on electron wave vector $\mathbf{k}$. The optical
transition of QUIP can be enhanced at $+\mathbf{k}$ but vanishes at
$-\mathbf{k}$, \textit{or} \textit{vise versa}. Therefore it is possible to
transfer the angular momentum between the photon and the electron-hole pair at
a specific $\mathbf{k}$ point, while it is forbidden at the opposite
$-\mathbf{k}$ point. Thus, the FR of QUIP can be used to detect the spin
population at opposite $\pm \mathbf{k}$ points independently, i.e., the pure
spin current case, by adjusting the external parameter such as the relative
phase between the $\omega$ and $2\omega$ excitations.

\begin{figure}[b]
\includegraphics[width=0.85\columnwidth,clip]{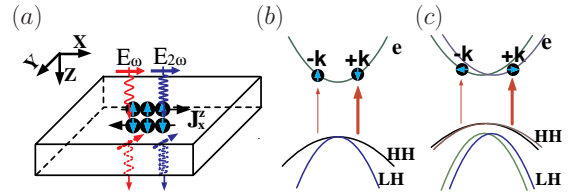} \caption{(color
online) Schematic illustration of two parallel-linearly polarized beams
propagating in semiconductor structures in the presence of a uniform pure spin
current ($a$). The band structure of a 3D sample without the SOI ($b$) and 2D
quantum well with the SOIs ($c$). The thickness of the red vertical arrows
indicates the strength of the transitions of QUIP.}%
\label{fig1}%
\end{figure}

Supposing there is a pure spin current in semiconductor structures, we study
the FR of a two-photon ($\omega$ and $2\omega$) quantum interference process.
First, instead of perturbation theory, we calculate the transition rate for
the one- and two- photon absorptions using the Volkov-type wave function which
describes the electron state under an optical field of arbitrary intensity
\cite{Volkov}. Then, we give the analytical expression of the FR of QUIP in a
3D sample. The valence and conduction states can be expressed as a Volkov-type
dressed state
\begin{align}
\psi_{c,v}(\mathbf{k},\mathbf{r},t) &  =u_{c,v}(\mathbf{k},\mathbf{r}%
)\exp[i\mathbf{k\cdot r-}i\omega_{c,v}t\nonumber \\
&  +\frac{ie}{m_{c,v}}\int_{0}^{t}\mathbf{k\cdot A}(\tau)d\tau
],\  \  \  \  \label{volkov}%
\end{align}
where $c$ ($v$) refers to the conduction (valence) band, $u_{c,v}%
(\mathbf{k},\mathbf{r})$ are the band-edge Bloch wave functions,
$E_{c,v}(k)=\hbar \omega_{c,v}(k)$ the energy of the valence or conduction
band, $m_{c,v}$ the effective masses, and $\mathbf{A}$ is the total vector
potential $\mathbf{A=a}_{1}A_{1}\cos \left(  \omega t+\varphi_{\omega}\right)
+\mathbf{a}_{2}A_{2}\cos \left(  2\omega t+\varphi_{2\omega}\right)  $. The
excitation field satisfies the relation $\hbar \omega<E_{g}<\hbar2\omega
<E_{g}+\Delta_{0}$, so that the $\omega$ and $2\omega$ excitations generate
independent carriers through one- and two- photon absorption. The transition
rate is calculated using a S-matrix formalism \cite{Volkov}
\begin{equation}
S=-\frac{i}{\hbar}\int_{-\infty}^{\infty}dt\int d^{3}r\psi_{c}^{\ast
}(\mathbf{k},\mathbf{r},t)H_{int}\psi_{v}(\mathbf{k},\mathbf{r}%
,t).\label{Smatrix}%
\end{equation}
The transition rate is
\begin{align}
W(\mathbf{k}) &  =C\{ \left(  \frac{\eta_{1}}{2}\right)  ^{2}\left \vert
\mathbf{P}_{cv}\cdot \mathbf{a}_{1}\right \vert ^{2}A_{1}^{2}+\left \vert
\mathbf{P}_{cv}\cdot \mathbf{a}_{2}\right \vert ^{2}A_{2}^{2}%
\  \  \  \  \  \  \  \  \nonumber \\
&  +\left[  A_{1}A_{2}\frac{\eta_{1}}{2}\left(  \mathbf{P}_{cv}\cdot
\mathbf{a}_{1}\right)  ^{\ast}\left(  \mathbf{P}_{cv}\cdot \mathbf{a}%
_{2}\right)  e^{i\left(  2\varphi_{\omega}-\varphi_{2\omega}\right)
}+c.c.\right]  \},\label{trans_rate}%
\end{align}
where $C=2\pi \left(  \frac{e}{2\hbar m_{0}c}\right)  \delta \lbrack \omega
_{cv}(k)-2\omega](f_{v}-f_{c})$, $f_{v}$ ($f_{c}$) is the Fermi function of
valence (conduction) electrons, $\mathbf{P}_{cv}=\langle c|\hat{p}|v\rangle$,
$\eta_{1}=\frac{eA_{1}}{\omega cm_{cv}}\mathbf{k\cdot a}_{\mathbf{1}}$, and
$1/m_{cv}=1/m_{c}-1/m_{v}$. The term in the square brackets describes the
quantum interference between the one- and two-photon excitations. From Eq.
(\ref{trans_rate}), the quantum interference term depends sensitively on the
electron wave vector $\mathbf{k}$, the electron spin orientation, the
polarization, and the relative phase of the pulses. The transition rate can be
strongly asymmetric for the transitions at $\pm \mathbf{k}$, e.g.,
$W(-\mathbf{k})=0$ while $W(+\mathbf{k})\neq0$ by adjusting the polarization
and the relative phase of the pulses.

The electron and hole states in Eq.(2) can be obtained from the single-band
and the multiband Luttinger-Kohn (LK) effective-mass Hamiltonian,
respectively. \cite{Luttinger} The LK Hamiltonian reads%
\begin{equation}
H_{h}(z_{h},\rho)=\frac{\hbar^{2}k^{2}}{2m_{0}}\left(
\begin{array}
[c]{cccc}%
H_{\mathit{hh}} & L & M & 0\\
L^{\ast} & H_{\mathit{lh}} & 0 & M\\
M^{\ast} & 0 & H_{\mathit{lh}} & -L\\
0 & M^{\ast} & -L^{\ast} & H_{\mathit{hh}}%
\end{array}
\right)  , \label{lutting}%
\end{equation}
where:

$H_{\mathit{hh}}=-\gamma_{1}-\gamma_{2}(\sin^{2}\theta_{e}-2\cos^{2}\theta
_{e}),$

$H_{\mathit{lh}}=-\gamma_{1}+\gamma_{2}(\sin^{2}\theta_{e}-2\cos^{2}\theta
_{e}),$

$M=-\sqrt{3}\gamma_{2}\sin^{2}\theta_{e}e^{-2i\varphi_{e}}, $

$L=i2\sqrt{3}\gamma_{3}\sin \theta_{e}\cos \theta_{e}e^{-i\varphi_{e}}$,\newline
where $\mathbf{k}=(k,\theta_{e},\varphi_{e})$, $\gamma_{1}$, $\gamma_{2}$, and
$\gamma_{3}$ are the Luttinger parameters. The energy dispersions of valence
subbands in the isotropic approximation ${\gamma_{2}=\gamma_{3}}$ are $E_{\pm
}=\frac{\hbar^{2}k^{2}}{2m_{0}}[-\gamma_{1}\pm(R_{h}^{2}+\left \vert
L\right \vert ^{2}+\left \vert M\right \vert ^{2})^{1/2}]$, and $R_{h}%
=-\gamma_{2}(\sin^{2}\theta_{e}-2\cos^{2}\theta_{e})$. The eigenstates of the
heavy-hole (\textit{hh}) and light-hole (\textit{lh}) bands are $|\mathit{hh}%
_{+}\rangle=\frac{1}{c_{h}}(R_{h}+E_{+}^{\prime},L^{\ast},M^{\ast},0)^{T}$,
$|\mathit{hh}_{-}\rangle=\frac{1}{c_{h}}(0,M,-L,R_{h}+E_{+}^{\prime})^{T}$,
and $|\mathit{lh}_{+}\rangle=\frac{1}{c_{h}}(L,-R_{h}+E_{-}^{\prime}%
,0,M^{\ast})^{T}$, $|\mathit{lh}_{-}\rangle=\frac{1}{c_{h}}(M,0,-R_{h}%
+E_{-}^{\prime},-L^{\ast})^{T}$, respectively, where $c_{h}$ is normalization
constant, and $E_{\pm}^{\prime}=\pm \sqrt{\left \vert M\right \vert
^{2}+\left \vert L\right \vert ^{2}+R_{h}^{2}}$ is the energy difference between
the \textit{hh} and \textit{lh} bands.

As shown schematically in Fig. \ref{fig1}($a$), the $\omega$ and $2\omega$
two-color optical fields are linearly parallel polarized along the x axis and
propagate along +z, $\mathbf{E}(\mathbf{r},t)=E_{\omega}e^{i(\omega
t-\mathbf{k_{\omega}\cdot r}+\varphi_{\omega})}\hat{e}_{\omega}+E_{2\omega
}e^{i(2\omega t-\mathbf{k_{2\omega}\cdot r}+\varphi_{2\omega})}\hat
{e}_{2\omega}$, and $\hat{e}_{\omega}$ and $\hat{e}_{2\omega}$ are the unit
polarization vectors. The propagation directions, $\mathbf{k_{\omega}}$ and
$\mathbf{k_{2\omega}}$ are both along $+\hat{z}$. Using Eq. (\ref{trans_rate}%
), the difference of the refractive index for the right- and left- circular
polarized lights in the presence of the pure spin current can be written as
\begin{align}
N_{+}-N_{-} \propto \underset{\mathbf{k}}{\sum}W_{+}(+\mathbf{k})+W_{+}%
(-\mathbf{k})-W_{-}(+\mathbf{k})-W_{-}(-\mathbf{k})\nonumber \\
=\underset{\mathbf{k}}{\sum}C_{0}[C_{i1}(f_{u}+f_{d})\cos(2\varphi_{\omega
}-\varphi_{2\omega})+C_{i2}\sin(2\varphi_{\omega}-\varphi_{2\omega
})\  \nonumber \\
\times \sin(2\varphi_{e})(f_{d}-f_{u})+C_{l}(f_{u}-f_{d})]\times \delta \left[
\omega_{cv}(k)-2\omega \right]  , \label{deps}%
\end{align}
where

$C_{0}=2\pi \frac{2}{\sqrt{2}c_{h}^{2}}\left(  \frac{e}{\hbar2m_{0}c}\right)
^{2}P^{2}$,

$C_{i1}=\frac{\eta_{1}^{\prime}}{2}A_{1}A_{2}k\sin \theta_{e}\cos \varphi_{e}%
\Re_{h,l}$,

$C_{i2}=\frac{\eta_{1}^{\prime}}{2}A_{1}A_{2}k\sin \theta_{e}\cos \varphi_{e}%
\Im_{l,h}\operatorname{Re}(M)/\sqrt{3}$,

$C_{l}=A_{2}^{2}\Re_{h,l}$,\newline where $\Re_{h}=[(R_{h}+E_{+}^{\prime}%
)^{2}-(|M|^{2}+|L|^{2})/3]$ and $\Im_{h}=2(R_{h}+E_{+}^{\prime})$ denote the
contributions from the transition from the \textit{hh} band to the conduction
band, and $\Re_{l}=[|M|^{2}+|L|^{2}-(-R_{h}+E_{-}^{\prime})^{2}/3]$ and
$\Im_{l}=2(-R_{h}+E_{-}^{\prime})$ for the transition from the \textit{lh}
band to the conduction band. $W_{+}(\pm \mathbf{k})$ ($W_{-}(\pm \mathbf{k})$)
denotes the transition rate for the left- (right-) circularly polarized
$2\omega$ optical field at $\pm \mathbf{k}$, and $f_{u}$ ($f_{d}$) is the
occupation number of spin-up (spin-down) electrons in the conduction band at
$+\mathbf{k}$ ($-\mathbf{k}$), $P^{2}=\left \vert \left \langle S\right \vert
P_{x}\left \vert X\right \rangle \right \vert ^{2}$, $\eta_{1}^{\prime}%
=\frac{eA_{1}}{\omega cm_{cv}}$. Thus, the FR rotation angle of the $2\omega$
pulse per unit length is given by $\theta_{F}(\omega)=\frac{\omega}%
{c}\operatorname{Re}(N_{+}-N_{-})$ \cite{shina}.

The first two terms in Eq. (\ref{deps}) represent the FR of QUIP arising from
the quantum interference between one- and two- photon absorptions. The
interference terms are linearly proportional to the wave vector $\mathbf{k}$
(see Eq. (\ref{trans_rate})), and consequently, FR is proportional to spin
current density $\mathbf{J_{i}^{z}}=\frac{1}{2}\mathbf{\langle \sigma_{z}%
v_{i}+v_{i}\sigma_{z}\rangle}\sim \underset{k}{\sum}[f_{u}(\mathbf{k}%
)+f_{d}(\mathbf{-k})]\mathbf{k},i,j=x,y,z$. \emph{Therefore}, the FR of QUIP
can detect the spin current directly. The third term in Eq. (\ref{deps}) is
the single-photon transition term corresponding to the conventional FR, which
is independent of the electron wave vector $\mathbf{k}$. For the pure spin
current, i.e., $f_{u}(\mathbf{k})=f_{d}(-\mathbf{k})$, the conventional FR
vanishes since the total spin of the pure spin current vanishes. Thus, the
conventional FR cannot detect the spin current directly if we neglect the
negligible small wave vector of light. The transition rates of one- and
two-photon absorptions can be set equal by tuning the intensity of laser
beams, i.e., $\eta_{1}A_{1}/2\simeq A_{2}$. The FR of QUIP in the detection of
pure spin current is on the same order as the conventional FR.

\begin{figure}[t]
\includegraphics[width=0.95\columnwidth,clip]{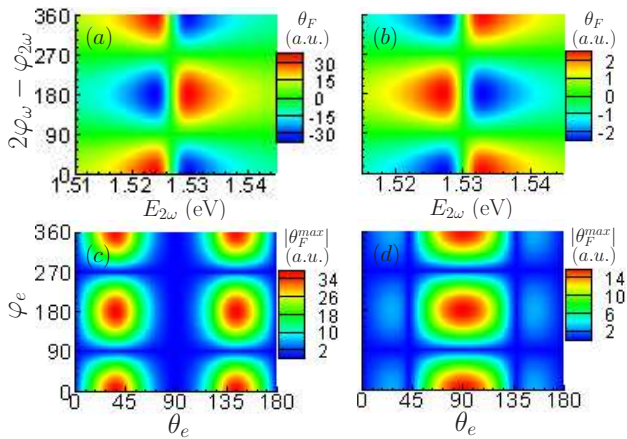}\caption{(color
online). Contour plot of FR of QUIP ($\theta_{F}$) as function of the energy
of the $2\omega$ optical field and relative phase of the lights for
$\theta_{e}=30^{\circ}$ for the \textit{hh}-conduction band [($a$)] and the
\textit{lh}-conduction band [($b$)] transition processes. ($c$) and ($d$) show
the contour plot of the QUIP FR ($|\theta_{F}^{max}|$) as function of polar
angle $\theta_{e}$ and $\varphi_{e}$ of elcetron wave vector for the
\textit{hh}-conduction band and for the \textit{lh}-conduction band transition
processes at $2\varphi_{\omega}-\varphi_{2\omega}=0^{\circ}$, respectively. }%
\label{fig2}%
\end{figure}

In order to detect the pure spin current $J_{j}^{i}(i,j=x,y,z)$ in
semiconductor structures, the magnitude and direction of electron velocity and
the electron spin orientation are necessary. Figures \ref{fig2} (a) and (b)
show the FR of QUIP as function of the energy and the relative phase between
the $\omega$ and $2\omega$ optical fields in the presence of a pure spin
current in bulk GaAs material\cite{para}. The FR of QUIP reaches its maxima at
$2\varphi_{\omega}-\varphi_{2\omega}=0^{\circ}$, $180^{\circ}$, and
$360^{\circ}$, and vanishes at $2\varphi_{\omega}-\varphi_{2\omega}=90^{\circ
}$ and $270^{\circ}$. One can determine the spin-up and spin-down electron
population $f_{u,d}$ at a specific wavevector $\mathbf{k}$ by setting the
different phase differences $2\varphi_{\omega}-\varphi_{2\omega}$ [see Eq.
(\ref{deps})]. The FR of QUIP can also be used to determine the direction of
the spin current and the polarization of the laser pulse [see the $\eta
_{1}\sim \mathbf{k\cdot a}_{\mathbf{1}}$ term in Eq. (\ref{trans_rate})], e.g.,
the FR of QUIP vanishes at $\varphi_{e}=90^{\circ}$ or $\theta_{e}=0^{\circ}$.
Figures \ref{fig2} (c) and (d) describe the maximum of the FR of QUIP as a
function of the direction of electron wave vector, i.e., polar angles
$\theta_{e}$ and $\varphi_{e}$ for $2\varphi_{\omega}-\varphi_{2\omega
}=0^{\circ}$. Utilizing the relationship between the FR of QUIP and the
direction of the electron wave vector, the direction of the spin current can
be determined experimentally. The oscillating FR of QUIP arises from the
interplay between the propagation direction of the light and the direction of
the electron wave vector, and the mixing of heavy-hole and light-hole states
[see Eq. (\ref{deps}) and $\Re_{h,l}$]. The hole mixing effect shortens the
oscillation period of the FR of QUIP from $2\pi$ to $\pi$.

Comparing the left panel with right panel of Figs. \ref{fig2}, the
contribution to the FR of QUIP from the \textit{hh}$_{\pm}$-\textit{c} band
transition is different from that from the \textit{lh}$_{\pm}$-\textit{c} band
transition due to the distinct energy dispersions and the Bloch band-edge wave
functions. The distinct energy dispersions make the optical transition occur
at different wavevectors for the \textit{hh}- and \textit{lh}-c band
transitions, while the Bloch band-edge wave functions of the \textit{hh} and
\textit{lh} bands determine the strengths of the corresponding transitions,
i.e., the magnitude of the FR of QUIP.

\begin{figure}[t]
\includegraphics[width=0.95\columnwidth,clip]{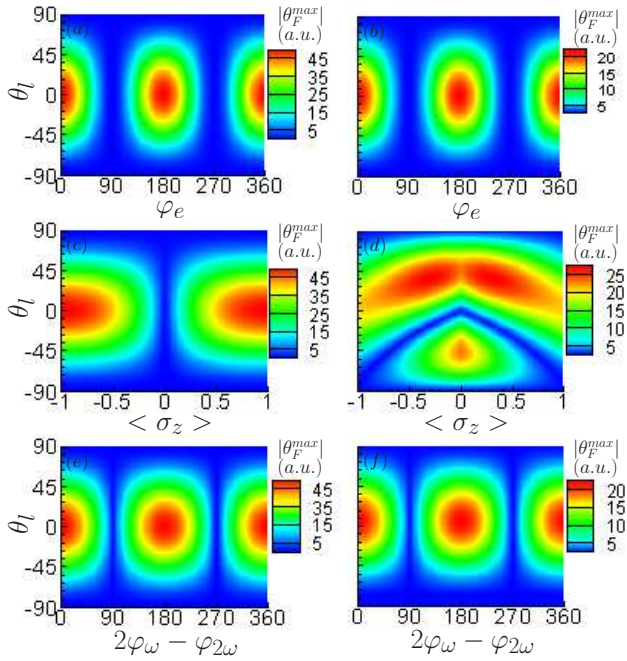}\caption{(color
online). Contour plot of FR of QUIP as a function of the direction of electron
wave vector and the propagation direction of pulses ($a$ and $b$), the spin
polarization direction and propagation direction of pulses ($c$ and $d$), the
relative phase and propagation direction of pulses ($e$ and $f$) for the
\textit{hh}-c band and \textit{lh}-c band transition processes.}%
\label{fig3}%
\end{figure}

Now we turn to discuss the detection of pure spin current in the 2D GaAs
quantum well\cite{para} in the presence of the spin-orbit interactions (SOIs)
[see Fig. \ref{fig1}(c)], i.e., Dresselhaus SOI (DSOI) and Rashba SOI (RSOI).
The difference between the 3D and 2D samples is the lifted degeneracy of the
\textit{hh} and \textit{lh} bands at $\Gamma$ point and the spin degeneracy
caused by the SOIs. We can detect the spin orientation and direction of the
pure spin current in the two-dimensional case. Likewise, the FR of QUIP also
vanishes when the polarization vector $\mathbf{a}_{1}$ is perpendicular to the
direction of the pure spin current. Figs. \ref{fig3}(a) and (b) depict the FR
of QUIP as a function of the direction of the electron wave vector
$\varphi_{e}$ and the propagation direction of pulse $\theta_{l}$ for an
out-of-plane oriented spin current $J_{\varphi_{e}}^{z}$. Figs. \ref{fig3}(c)
and (d) show the FR of QUIP as a function of the electron spin orientation and
the propagation direction of the laser pulse. When the spin orientation is
along the $z$ axis, the FR of QUIP from the \textit{hh}-c band transition
increases since the $|X\rangle$ component of the \textit{hh} band is crucial
for the FR of QUIP. A strongly anisotropic FR can be found in Fig.
\ref{fig3}(d) as function of the propagation direction of laser pulses and the
spin orientation $<\sigma_{z}>$, which is caused by the anisotropic transition
matrix elements $\langle c|\mathbf{p}|v\rangle \sim-\frac{1}{\sqrt{6}}\langle
S|\cos \theta_{l}p_{x}|X\rangle+\sqrt{\frac{2}{3}}\langle S|\sin \theta_{l}%
p_{z}|Z\rangle$. Thus, we can obtain information about the spin orientation of
spin currents by altering the propagation direction of laser pulses. We find
that the transition between the \textit{lh-c} bands leads to a larger FR than
that of the \textit{hh}-c transition since the FR of QUIP for in-plane
oriented spin current is dominantly determined by the $|Z\rangle$ component in
the hole states. The $|Z\rangle$ component in the band-edge Bloch wave
function of the light hole band $|lh\rangle=(1/\sqrt{6})[|X\rangle \pm
i|Y\rangle \otimes|\downarrow \uparrow \rangle-2|Z\rangle \otimes|\uparrow
\downarrow \rangle]$ is stronger than that in the heavy hole band
$|hh\rangle=(1/\sqrt{2})(|X\rangle \pm i|Y\rangle)\otimes|\uparrow
\downarrow \rangle$ in which the $|Z\rangle$ component primarily comes from the
hole mixing effect which is strong for large wave vectors. As in 3D sample,
the relative phase also can be used to tune the FR of QUIP (see Fig.
\ref{fig3}(e) and (f)), which shows a cosine function with the relative phase.

Finally, we can also estimate the magnitude of FR of QUIP in realistic
semiconductor systems. For the optical quantum interference injection
\cite{Stevens} in GaAs QWs, the carrier density is about $10^{17}cm^{-3}$, and
the corresponding FR of QUIP from our calculations is about $20mrad/\mu m$. In
the recent experiment\cite{stern}, the spin Hall conductivity is about
$3\Omega^{-1}m^{-1}/|e|$ at the electric field $E=36.2mV/\mu m$, the
corresponding spin-up (spin-down) electron density is about $10^{13}cm^{-3}$,
and the FR of QUIP for this case is about $2\mu rad/\mu m$, which is not
difficult to detect using the current FR\ technique. The ultrafast lasers can
control quantum system on the femtosecond time scale which is shorter than the
electron spin lifetimes and electron momentum relaxation times (about
0.19\textit{ps}-1.2\textit{ps}) \cite{oudar}, especially in the n-type layers,
a long spin lifetime ($10^{3}-10^{4}$\textit{ps}) can be achieved
\cite{Kato,kikka}.

In summary, we have suggested a new scheme for the direct optical detection of
a uniform pure spin current utilizing the two-color momentum-resolved FR of
QUIP in bulk semiconductors and two-dimensional semiconductor quantum well
structures. By adjusting the incident angle, the polarization direction, the
energy, and the relative phase between the $\omega$ and $2\omega$ laser
pulses, detailed information about the spin current can be detected directly.
This scheme may also be important for distinguishing the ESHE and ISHE from
the different dependence of the pure spin current on the crystal orientation.
Our scheme can also be used to detect the spin polarization of the charge current.

This work was supported by the NSFC Grant No. 60525405 and the
knowledge innovation project of CAS. K.C. would like to appreciate
R. B. Liu for discussions.


\begin{thebibliography}{99}
\bibitem {para}The following parameters of GaAs are used in our numerical
calculation: $m_{e}=0.0665m_{0}$ ($m_{0}$ is the free electron mass),
$\gamma_{1}=6.85$, $\gamma_{2}=2.1$, and the band gap $E_{g}=1.52eV$. The
band-gap mismatch between $GaAs$ and $Al_{x}Ga_{1-x}As$ $\bigtriangleup
E_{g}=1.155x+0.37x^{2}$, $x=0.35$, $\bigtriangleup E_{e}/\bigtriangleup
E_{v}=60/40$, DSOI and RSOI parameters are taken from Ref. \cite{yang,
pala,rashba}: $E_{z}=60kV/cm$, $\langle \alpha E_{z}\rangle=1meV\cdot nm$,
$\langle \beta E_{z}\rangle=7.5meV\cdot nm$, $\alpha_{c}^{\ast}=35eV{\mathring
{A}}^{3}$, $\alpha_{v}=-39eV{\mathring{A}}^{3}$, $\delta \alpha_{v}%
=-35eV{\mathring{A}}^{3}$.
\end{thebibliography}
\end{document}